\RequirePackage{fix-cm}
\documentclass[twocolumn,epjc3]{svjour3}  
\smartqed  
\RequirePackage{graphicx}

\journalname{Eur. Phys. J. D}

\usepackage{graphicx}
\usepackage{bm}
\usepackage{hyperref}
\usepackage{amsmath}
\usepackage{braket}
\usepackage{dsfont}
\usepackage{siunitx}
\usepackage{color}

\newcommand{\ee}{\mathrm{e}}
\newcommand{\ii}{\mathrm{i}}

\usepackage{cite}
\begin{document}

\title{Edge-state influence on high-order harmonic generation in topological nanoribbons}
\author{Christoph J\"ur{\ss}\thanksref{e1,addr1} \and Dieter Bauer\thanksref{e2,addr1}}

\thankstext{e1}{e-mail: christoph.juerss@uni-rostock.de}
\thankstext{e2}{e-mail: dieter.bauer@uni-rostock.de}

\institute{Institute of Physics, University of Rostock, 18051 Rostock, Germany\label{addr1}}
\date{Received: date / Revised version: date}
%
\abstractdc{
The high-order harmonic generation in finite topological nanoribbons is investigated using a tight-binding description. The ribbons consist of hexagons and are almost one-dimensional. Two edge states emerge at the short edges of the ribbon after a gap closure between valence and conduction band, indicating a topological phase transition. The energies of those edge states as functions of  the tight-binding parameters display  crossings and avoided crossings, which influence the high-harmonic spectra. 
} 

\sloppy

\maketitle
\section{Introduction} 
	
	    Topological insulators are a special kind of solid state material that is an electrical insulator in its bulk but conducting on its edges or surfaces. The edge or surface states are protected against perturbations \cite{topinsRevModPhys.82.3045}. 
	    
	    Recent studies show that the topological phase of a solid can have a huge influence on the generation of high-order harmonic radiation.  In fact, the topological phase might affect the harmonic yield by several orders of magnitude \cite{bauer_high-harmonic_2018,DrueekeBauer2019,JuerssBauer2019}, flip the helicity of the emitted photons \cite{Silva2019,juerss2020helicity,Moos_HHG_solids} or introduce  circular dichroism  \cite{chacon_observing_2018}. In three-dimensional topological insulators the harmonic yield of bulk and surface states show a different dependency on the ellipticity of the laser field \cite{Baykusheva_2021}.
	    
	    It is known that the high-harmonic generation (HHG) in solids in general carries information about the static and dynamic properties of the solid \cite{VampaPhysRevLett.115.193603,Hohenleutner2015,Luu2015,vampa_merge_2017,You2017,Baudisch2018}.	   In this work, we investigate hexagonal nanoribbons that are almost one-dimensional. The systems are described by a tight-binding model where hopping between nearest neighbors are allowed, thus describing  graphene ribbons. The HHG in graphene was studied previously, for example in Refs.\ \cite{PhysRevB.95.035405,Chizhova_2017,Yoshikawa2017,Hafez2018}. Adding an alternating on-site potential because of different atomic elements such as in hexagonal boron nitride (h-BN), for instance, the  sublattice symmetry is broken.   HHG in h-BN has been studied as well, e.g.\ in Refs.\ \cite{Tancogne-Dejeaneaao5207,LeBreton_2018,Yue2020}.  The dependence of HHG on the on-site potential for hexagonal ribbons was studied in \cite{juerss_bauer2021highorder,drueeke_bauer2021highharmonic}. 
	    
	    With a broken time-reversal symmetry, the system might become topologically nontrivial. This can be achieved by including a complex hopping between next-nearest neighbors as in the Haldane-model \cite{Haldane_1988}. The Haldane-model in the context of HHG  was studied in Refs.\ \cite{Silva2019,chacon_observing_2018, juerss2020helicity}.
	    
	    In this paper, we examine how the edge states of Haldane nanoribbons  influence the emission of high-order harmonics. Topological nanoribbons were studied without an external field in Ref.\ \cite{Hao_Haldane_ribbons}. Although edge states are only present in finite systems the bulk-boundary correspondence \cite{topinsshortcourse} tells that a nonvanishing difference between the topological invariants of the bands for the bulk  imply the presence of edge states in the respective finite system. The question then is which topological effects in HHG spectra are due to bulk already and which require the explicit presence of edge states. An example system where the explicit presence of edge states is necessary to see any topological effect in HHG spectra is the one-dimensional Su-Schrieffer-Heeger chain \cite{Moos_HHG_solids}. In 2D systems such as the Haldane model, on the other hand, one can observe helicity flips already for bulk only  \cite{Silva2019,Moos_HHG_solids}.

	    The outline of the paper is as follows.
	    In Sec. \ref{sec:theory}, we  summarize the theoretical methods used in this work. In Sec. \ref{sec:res_static}, the properties of the static system are explained, with a focus on the edge states. The HHG of Haldane nanoribbons is discussed in Sec. \ref{sec:res_HHG}. If not stated otherwise, atomic units (a.u., $\hbar=|e|=m_e=4\pi\epsilon_0=1$) are used throughout this paper.
	
	\section{Theory}\label{sec:theory}
        
        \begin{figure} 
			\includegraphics[width = \columnwidth]{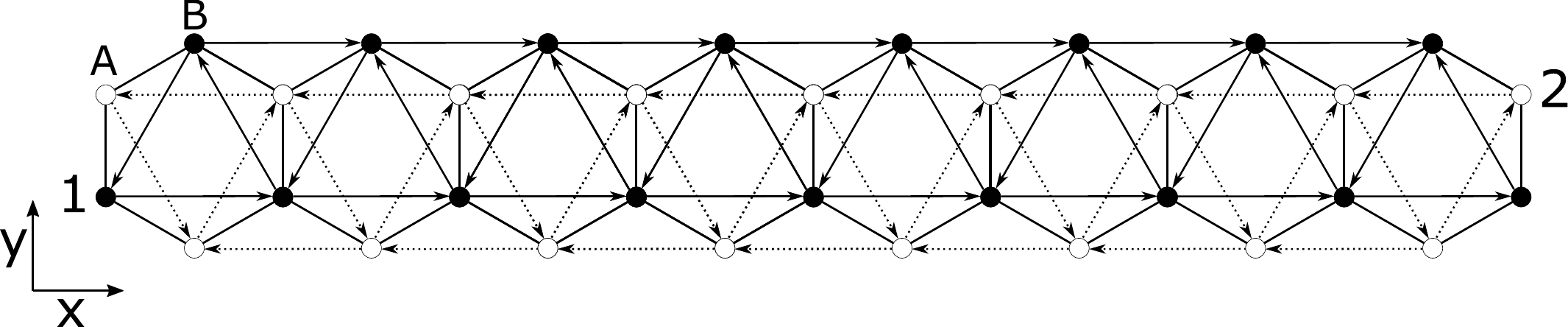}
			\caption{\label{fig1} Sketch of a nanoribbon with zig-zag edges comprising eight hexagons. The circles indicate the atomic sites. The on-site potential on sites with an unfilled circle is given by $M$ (sublattice A),  for the filled circles it is given by $-M$ (sublattice site B). Lines (without arrow) indicate the hopping between nearest neighbors (amplitude $t_1$). The arrows indicate the next-nearest neighbor hopping with amplitude  $\ii t_2$ in the direction of the arrow (and $-\ii t_2$ in the opposite direction). We label the lower left edge site `1' and the upper right  edge site `2'.}
		\end{figure}
        
        The systems that are investigated in this work are hexagonal ribbons with zig-zag edges as sketched in Fig.\ \ref{fig1}.  The simulated ribbons consist of 30 hexagons though (not only eight, as sketched in Fig.\ \ref{fig1}). A tight-binding approximation is used. The circles in Fig. \ref{fig1}  indicate the sites with an on-site potential $M$ ($-M$) for the unfilled (filled) circles, corresponding to the
         two sublattice sites A (unfilled) and B (filled). Lines without arrows indicate hopping between the nearest neighboring sites with an
        amplitude  $t_1 \in \mathds{R}$. The arrows indicate a complex next-nearest neighbor hopping with amplitude $\ii t_2$ (with $t_2 \in \mathds{R}$) along the arrows (and $-\ii t_2$ in the opposite direction). The complex next-nearest neighbor hopping breaks the time-reversal symmetry, making the system topologically nontrivial for sufficiently large $t_2$ \cite{Haldane_1988}.

		\subsection{Static system}\label{sec:static}
		    
		    The theoretical description of the topological ribbons is almost the same as in Ref.\ \cite{juerss2020helicity}, with the  difference that periodic boundary conditions were assumed there. As a consequence,   the hopping elements from the left to the right edge of the ribbon are missing in the present work.
		    
		    The Hamiltonian describing the electrons on the  ribbon reads in tight-binding approximation
		    \begin{align}\label{eq:t_indep_H}
		        \hat{H}_0 &= t_1\sum_{<i,j>}
		        \left( \ket{j}\bra{i} + \mathrm{h.c.}\right) \, +\, \ii\, t_2\sum_{\ll i,j\gg}
		        \left( \ket{j}\bra{i} - \mathrm{h.c.}\right)\nonumber\\
		        &+ M\left(\sum_{i\in A} \ket{i}\bra{i} - \sum_{i\in B} \ket{i}\bra{i}\right)
		    \end{align}
		    where the sums $\sum_{<i,j>}$ and $\sum_{\ll i,j\gg}$ run over all nearest and next-nearest neighboring sites $i$, $j$, respectively. The sums $\sum_{i\in A}$ ($\sum_{i\in B}$) include all sites on sublattice site A (B). The orbital $\ket{i}$ denotes the atomic orbital at site $i$. A general state reads 
		    \begin{equation}
		        \ket{\psi} = \sum_{i = 1}^N g_i\ket{i},
		    \end{equation}
		    where $N$ is the number of sites in the system. 
		    
		    The time-independent Schr\"odinger equation
		    \begin{equation}\label{eq:TISE}
		        \hat{H}_0\ket{\psi_l} = E_l\ket{\psi_l}
		    \end{equation}
		    is solved to obtain the eigenstates $\ket{\psi_l}$ with their respective energies $E_l$. The number of eigenstates 
		    is given by the number of sites $N$, i.e. $l = 0,1,2,...,N-1$. The labeling is such that the energies of the states increase with $l$, i.e., $E_0\leq E_1 \leq E_2 \leq ...\leq E_{N-1}$. Equation (\ref{eq:TISE}) is solved numerically by diagonalization of the Hamiltonian (\ref{eq:t_indep_H}). 
		    
		    The distance between nearest neighbors is set to $a=2.68$\,a.u.$\,\simeq\, 1.42$\,\AA\, and the hopping between them to $t_1 =- 0.1 \, ~$a.u.$\,\simeq \,2.7$\,eV, the parameters for graphene \cite{Cooper_2012}. The on-site potential $M$ and the next-nearest neighbor hopping amplitude $t_2$ are varied in this work.
		    
        \subsection{Coupling to an external field}\label{sec:coupling_field}
        
            The ribbons are coupled to an external field via velocity gauge, which translates to the Peierls substitution \cite{Peierls1933} in  tight-binding approximation. The gauge-invariant coupling of general tight-binding systems to external fields   was derived in Ref. \cite{Graf_1995}.
            
            The laser pulses are described by a  vector potential of the form 
		    \begin{align}
    		    \bm{A}(t) = A_0~\sin ^2\left(\frac{\omega_0 t}{2 n_{cyc}}\right)~\sin (\omega_0 t) \bm{e}_x,
    		\end{align}
    		for times  $0\leq t \leq 2\pi n_{cyc}/\omega_0$ (and zero otherwise). It is linearly polarized along the ribbon, that is, in $x$-direction. The number of cycles in the laser pulse is chosen   $n_{cyc} = 5$, the amplitude of the vector potential  $A_0 = 0.05$ (intensity  $\simeq 5\times 10^{9}~ \mathrm{Wcm}^{-2}$), and the angular frequency is $\omega_0 = 7.5\cdot 10^{-3}$ (i.e., wavelength $\lambda = \SI{6.1}{\micro\meter}$).
		    
		    We assume that all states with an energy smaller than $E = 0$ are occupied. Due to the symmetry of the energy spectrum, these are half of the states. Hence,  $\ket{\Psi_l(t)}$ with $l = 0,1,2,...,N/2-1$ are propagated in time,  starting from $\ket{\Psi_l(t=0)} = \ket{\psi_l}$.
		    
		    The total current is given by  
		    \begin{align}\label{eq:current}
			    \bm{J}(t) = \sum_{l}^{N/2-1}\bra{\Psi^l(t)}\hat{\bm{j}}(t)\ket{\Psi^l(t)},
		    \end{align}
		    where the current operator is given by \cite{Review_Transport}
    		\begin{align}\label{eq:current_operator}
    			\hat{\bm{j}}(t) = 	-\mathrm{i}\sum_{i,j}\left(\bm{r}_{i} - \bm{r}_{j} \right)\ket{i}\bra{i}\hat{H}(t)\ket{j}\bra{j},
    		\end{align}
		    with the positions $\bm{r}_{i,j}$ of sites $i, j$. The time-dependent Hamiltonian reads
		    \begin{align}
		        \bra{i}\hat{H}(t)\ket{j} = \bra{i}\hat{H}_0\ket{j}\ee^{-\ii \left(\bm{r}_{i} - \bm{r}_{j}\right)\cdot \bm{A}(t)}.
		    \end{align}
            Harmonic spectra are calculated from the two components of the current (\ref{eq:current}) via fast Fourier transformation,
		    \begin{align}
		        P_{\parallel,\perp}(\omega) = \mathrm{FFT}\left[\dot{J}_{x,y}(t)\right].
		    \end{align}
		     Here, $\parallel$ and $\perp$ denote the polarization direction of the emitted light with respect to the incoming field: parallel ($x$-direction) and perpendicular ($y$-direction), respectively.
		    The functions $\left|P_{\parallel,\perp}(\omega)\right|^2$ are proportional to the intensity of the emitted light \cite{Bandrauk2009,Baggesen_2011,bauer_computational_2017} polarized  in the respective direction. The phase difference
		    \begin{equation}
		        \Delta \phi = \mathrm{arg}\left(P_{\parallel}(\omega)P^*_{\perp}(\omega)\right) \label{eq:phasediff}
		    \end{equation}
            indicates the helicity of the emitted photons.

	\section{Results}\label{sec:results}
	
	\subsection{Static system}\label{sec:res_static}
	    \begin{figure} 
			\includegraphics[width = \columnwidth]{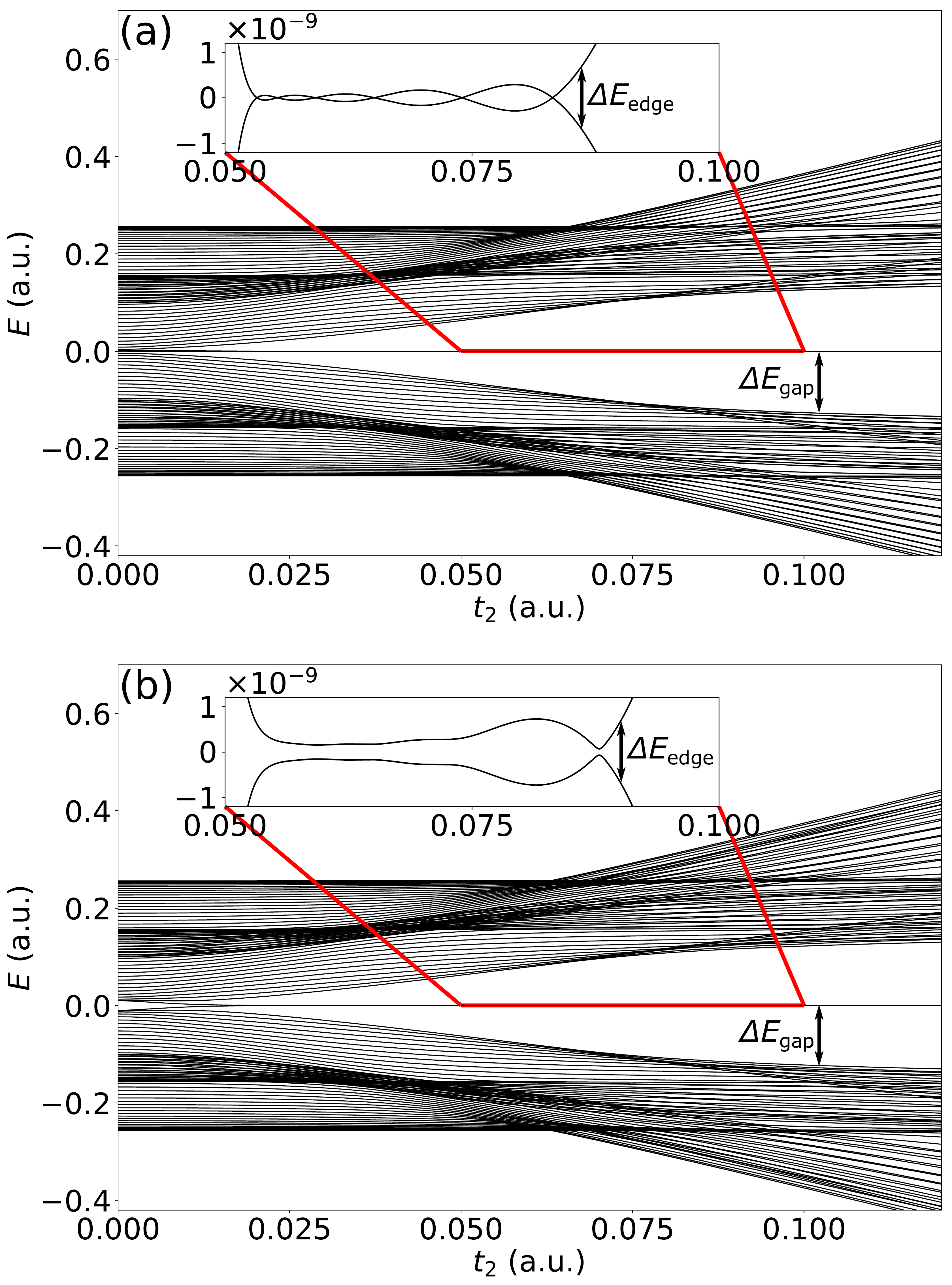}
			\caption{\label{fig2} Energies of the system for (a) $M = 0$ and (b) $M = 0.01$ as function of $t_2$. The insets show the evolution of the two states in a tiny energy interval  around $E= 0$.}
		\end{figure}
		
	     The number of atoms and eigenstates for the 30-hexagon long ribbons is $N = 122$. In Fig.\ \ref{fig2}, the energies of all states as function of the next-nearest neighbor hopping amplitude $t_2$ are shown for $M=0$ (Fig.\ \ref{fig2}a) and  $M = 0.01$ (Fig.\ \ref{fig2}b). 
	    For $M = 0$ (Fig. \ref{fig2}a) a band gap opens as $t_2$ increases so that the fully occupied valence band with $E<0$ and the empty conduction band with $E>0$ become well separated. In the middle of the gap, two states appear around an energy of $E = 0$. One of this state is occupied, the other one not.
	    For the system with an on-site potential of $M = 0,01$ (Fig. \ref{fig2}b) there is already a band gap for $t_2 = 0$ but without states in the middle. First, this band gap closes with increasing $t_2$ before it opens up again  for larger values. This band gap closure is an indication for a topological phase transition. In fact,  in the middle of the band gap two states appear when the band gap opens up again. We will call those two states edge states because their probability density is located on the edges of the chain, as we will show in Fig. \ref{fig4}. We define the energy difference $\Delta E_\mathrm{gap}$ as the energy difference between the valence band and the lowest edge state energy.  
	    
	    The insets in Figs. \ref{fig2}a,b are magnifications and  show the tiny energies of both edge states between $0.05 \leq t_2 \leq 0.01$. Their difference is defined as $\Delta E _\mathrm{edge}$. Surprisingly, the energies of these states do not just monotonically  converge to $E= 0$. For $M = 0$ they cross six times in the interval shown before their energies separate for larger $t_2$. For the finite on-site potential in Fig.\ \ref{fig2}b  the crossings turn into avoided crossings.
	    
	    \begin{figure} 
			\includegraphics[width = \columnwidth]{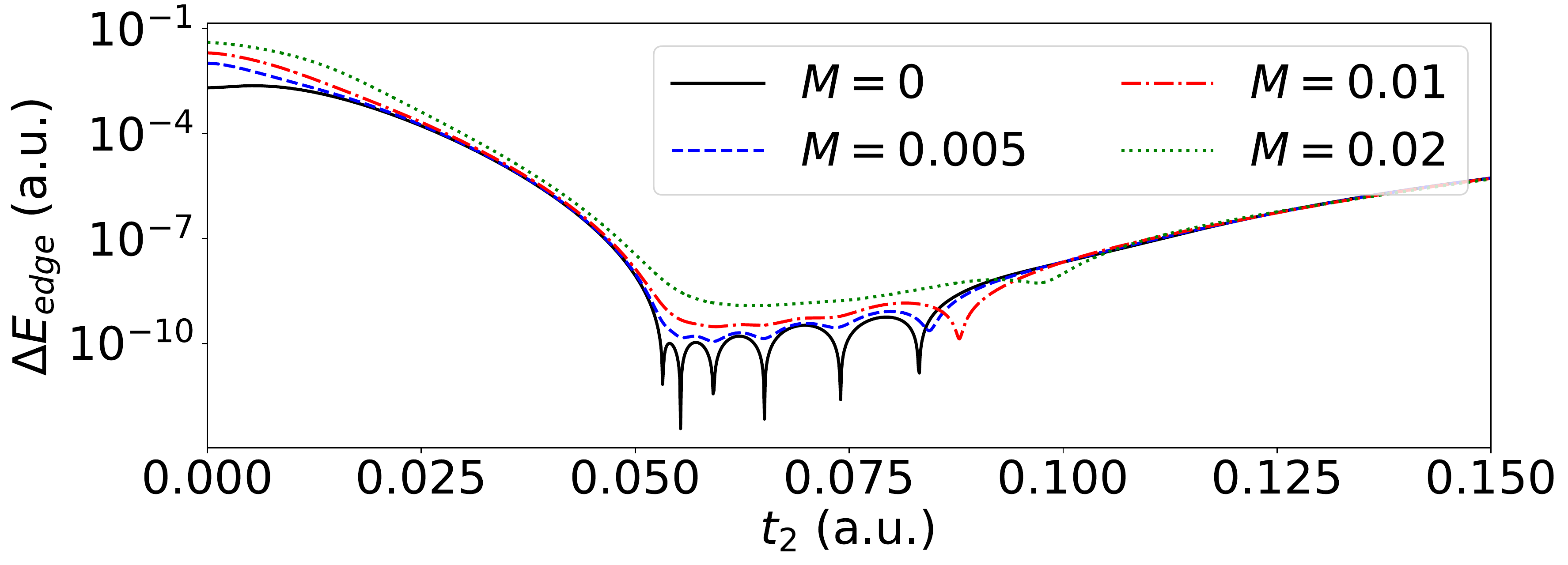}
			\caption{\label{fig3} Energy difference between the edge states as function of $t_2$ on a logarithmic scale for different on-site potentials $M$.}  
		\end{figure}

		In Fig.\ \ref{fig3}, the energy difference between the edge states $\Delta E_\mathrm{edge}$ is shown for different $M$ as function of $t_2$. It shows the crossings for $M = 0$ and that these crossings become avoided crossings for larger $M$. As the on-site potential increases further, the avoided crossings tend to smooth out. 
		For $M = 0.01$ and $M = 0.02$ there are mainly two local minima. The energy difference $\Delta E_\mathrm{edge}$ decreases with $t_2$ up to a local minimum at around $t_2 = 0.059$ for $M = 0.01$. The slope of $\Delta E_\mathrm{edge}$ following this local minimum is quite shallow, rendering it a flat local minimum. A much more localized minimum occurs ar $t_2 = 0.088$. Both minima are shifted towards larger $t_2$ as $M$ increases. The other crossings cannot be observed anymore as the on-site potential becomes larger.   
		
		\begin{figure} 
			\includegraphics[width = \columnwidth]{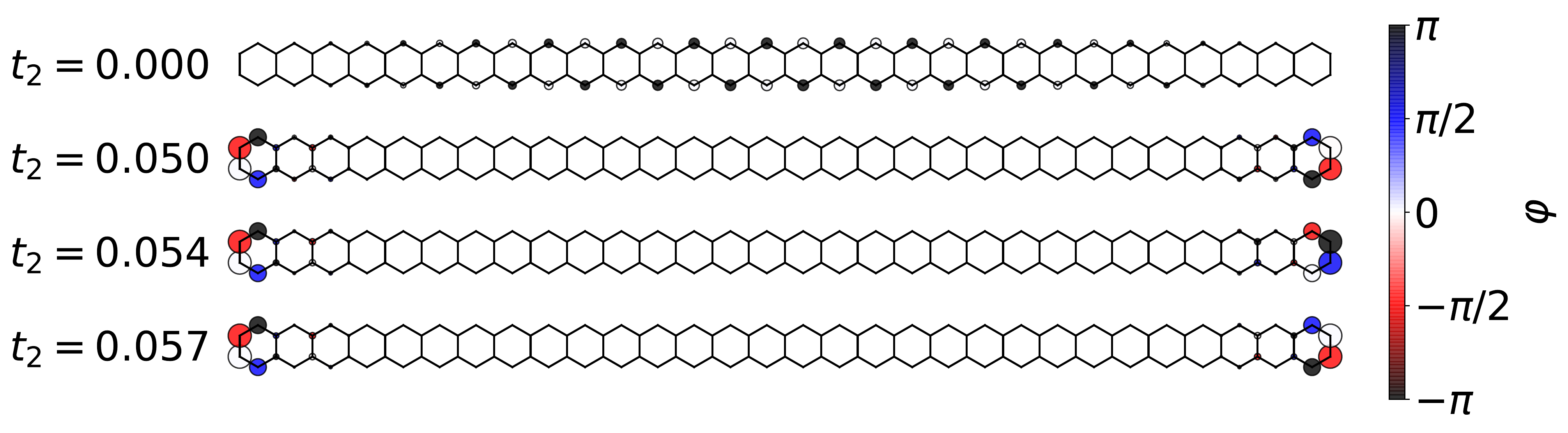}
			\caption{\label{fig4} Wave function of the highest occupied state (lowest state of the two edge states) for $M = 0$ and different $t_2$.  The size of the circles indicates the probability density. The phase of the wave function is indicated by  the color of the circles.}  
		\end{figure}
		
		\begin{figure} 
			\includegraphics[width = \columnwidth]{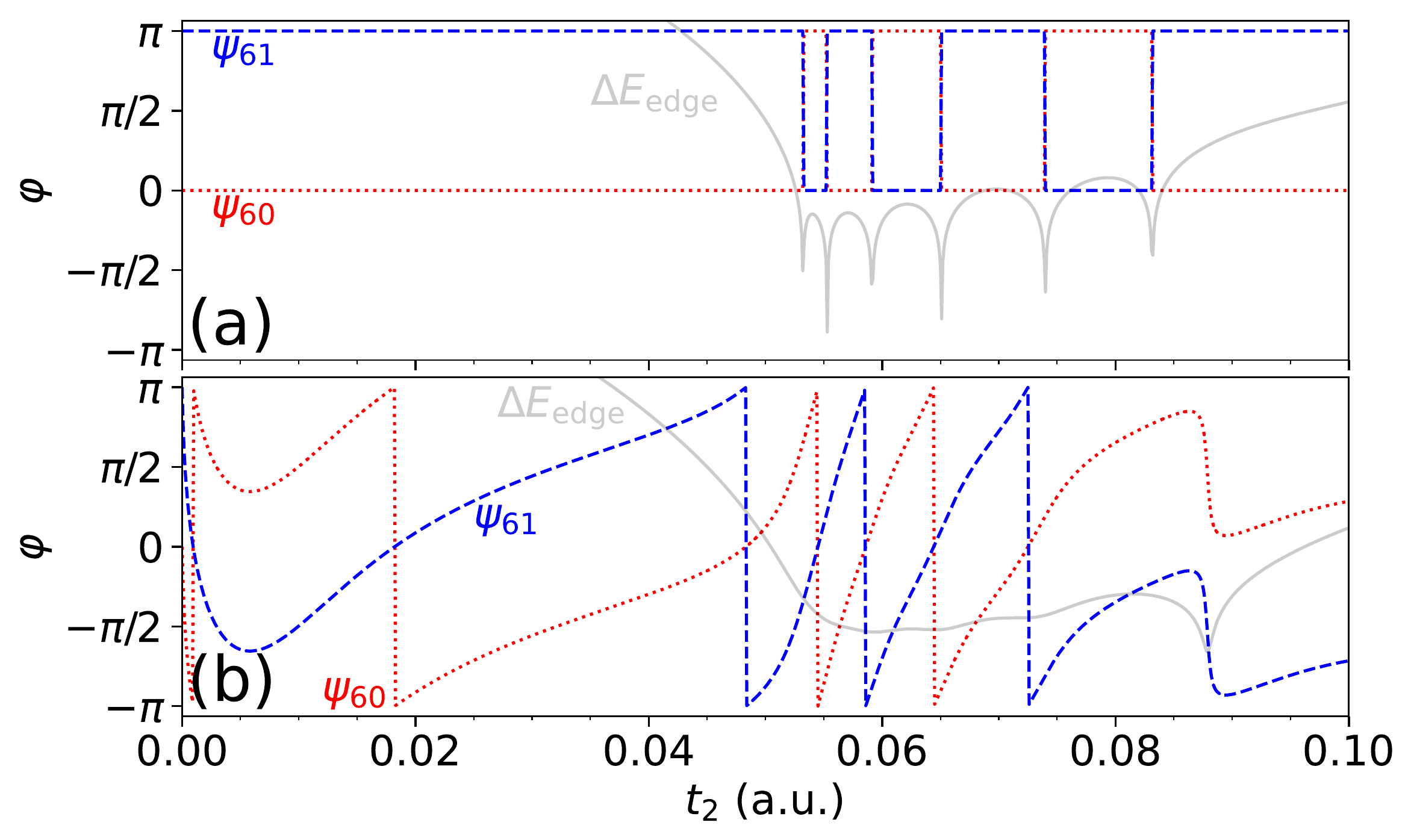}
			\caption{\label{fig5} Phase of the wave function of the initially highest occupied state ($\Psi_{60}$) and the lowest unoccupied state ($\Psi_{61}$) on site 2 (see Fig. \ref{fig1}) of the ribbon as function of $t_2$ for an on-site potential  $M = 0$ (a) and $M = 0.01$ (b).  The differences of the energies of both states $\Delta E_\mathrm{edge}$ on a logarithmic scale from Fig.\ \ref{fig3} are included in gray to show that phase jumps occur at (avoided) crossings.  }  
		\end{figure}

	    In Fig.\ \ref{fig4}, the wave function $\psi_{60}$ of the state within the band gap with the smaller energy  is shown. This is the highest, initially  occupied  state. The size of the sketched circles at the  lattice  site  scales with the probability density there. The phase of the wave function is indicated by the color of the circles. For ease of comparison, we use a phase convention for the initial states for which the phase at site 1 is zero (see Fig. \ref{fig1}). The wave functions are given for different $t_2$ but fixed on-site potential $M = 0$. For $t_2 = 0$ the probability density is equally located on the upper and lower edge. With increasing $t_2$, the electron probability moves towards the left and right edges. The same happens for the lowest unoccupied state $\psi_{61}$, whose phase is different but probability density is the same (not shown). Because of the dominant location of the electron at the edges we call these two states edge states. Clearly, for a system periodic  in $x$-direction these kind of states are absent  due to the absence of left and right edges.
	    
	    Crossings of the energies between both edge states occur at $t_2 = 0.053$ and $t_2 = 0.055$. For $t_2 = 0.05$ one can see a certain symmetry of the highest occupied state in Fig.\ \ref{fig4}. The phases of the wave function at the four leftmost sites, reading from top to bottom, is identical to the phases at the four rightmost sites but reading from bottom to top. The electron is mainly located on those eight sites. The wave function is symmetric under rotation by $180^\circ$ about an axis perpendicular to the $xy$-plane of the ribbon and through its center. For $t_2 = 0.054$ the energies of both edge states have crossed so that the occupied state should now have the properties of the (for lower $t_2$) unoccupied one, and the other way around. Indeed, the phases of the wave function on the right edge of state $\psi_{60}$ are now different. The state is not symmetric anymore under rotations by $180^\circ$. The phases at the four rightmost sites reading, from bottom to top, is identical to the phase at the four leftmost sites, read from top to bottom, plus $\pi$. This is indeed the symmetry of the other edge state. The next crossing appears at $t_2 = 0.055$. The symmetry of the highest occupied state for $t_2 = 0.057$ is now identical to the state for $t_2 = 0.05$, indicating that another crossing occurred.
	    
	    In order to identify the exchange of the edge states it is sufficient to look at the phases at, e.g.,  site 2 (see Fig. \ref{fig1}). In Fig.\ \ref{fig5}, the phase at this site for both edge states $\psi_{60}$ and $\psi_{61}$ is shown as function of $t_2$. In Fig.\ \ref{fig5}a, the phases for $M = 0$ are shown. The phases are constant for small $t_2$. For the highest occupied state $\psi_{60}$ the phase is $\varphi = 0$ and $\varphi = \pi$ for the lowest unoccupied state $\psi_{61}$. At each crossing the phases of both states change to the value of the other state, indicating that the properties of both states are exchanged each time their energies cross. In order to remind for which $t_2$  crossings occur,   the energy difference $\Delta E_\mathrm{edge}$  from Fig.\ \ref{fig3} is sketched in gray. In Fig.\ \ref{fig5}b, the same is plotted for an on-site potential of $M = 0.01$. From a local minimum at $t_2=0.0058$ on both phases increase up to $t_2 = 0.088$ where $\Delta E_\mathrm{edge}$ assumes a minimum (note that we plot phases modulo $2\pi$ within the interval $[-\pi,\pi)$ so that phases exceeding $\pi$ reenter at $-\pi$). In a narrow neighborhood around this value of $t_2$  both phases change by about $\pi$ in a continuous manner, which is characteristic of an  avoided crossing.  The properties of the two edge states also exchange in this case so that  the previously highest occupied state becomes the previously lowest unoccupied state and the other way around.

	   \subsection{High-harmonic generation}\label{sec:res_HHG}
	   
	   \begin{figure*}
        \includegraphics[width = \linewidth]{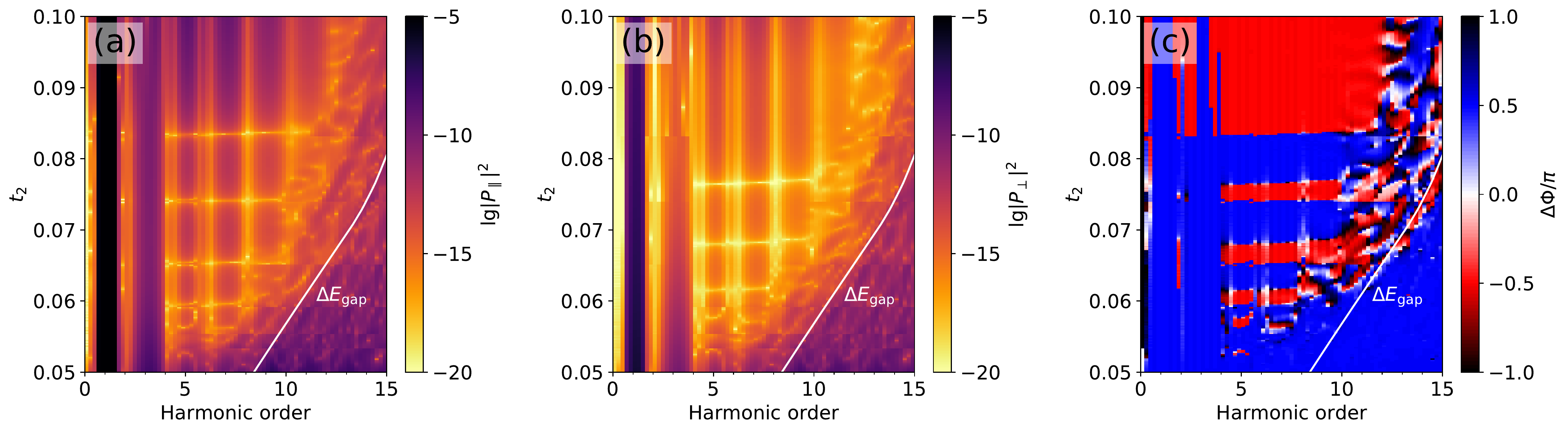}
        \caption{\label{fig6} High-order harmonic spectrum in (a) parallel and (b) perpendicular direction to the polarization of the external field, and (c) the corresponding phase difference between both components as a contour-plot as function of $t_2$. The on-site potential is $M = 0.0$. The white line $\Delta E_\mathrm{gap}$ is the gap between the valence band and the lowest edge state.}
        \end{figure*}
	   
	   \begin{figure*}
        \includegraphics[width = \linewidth]{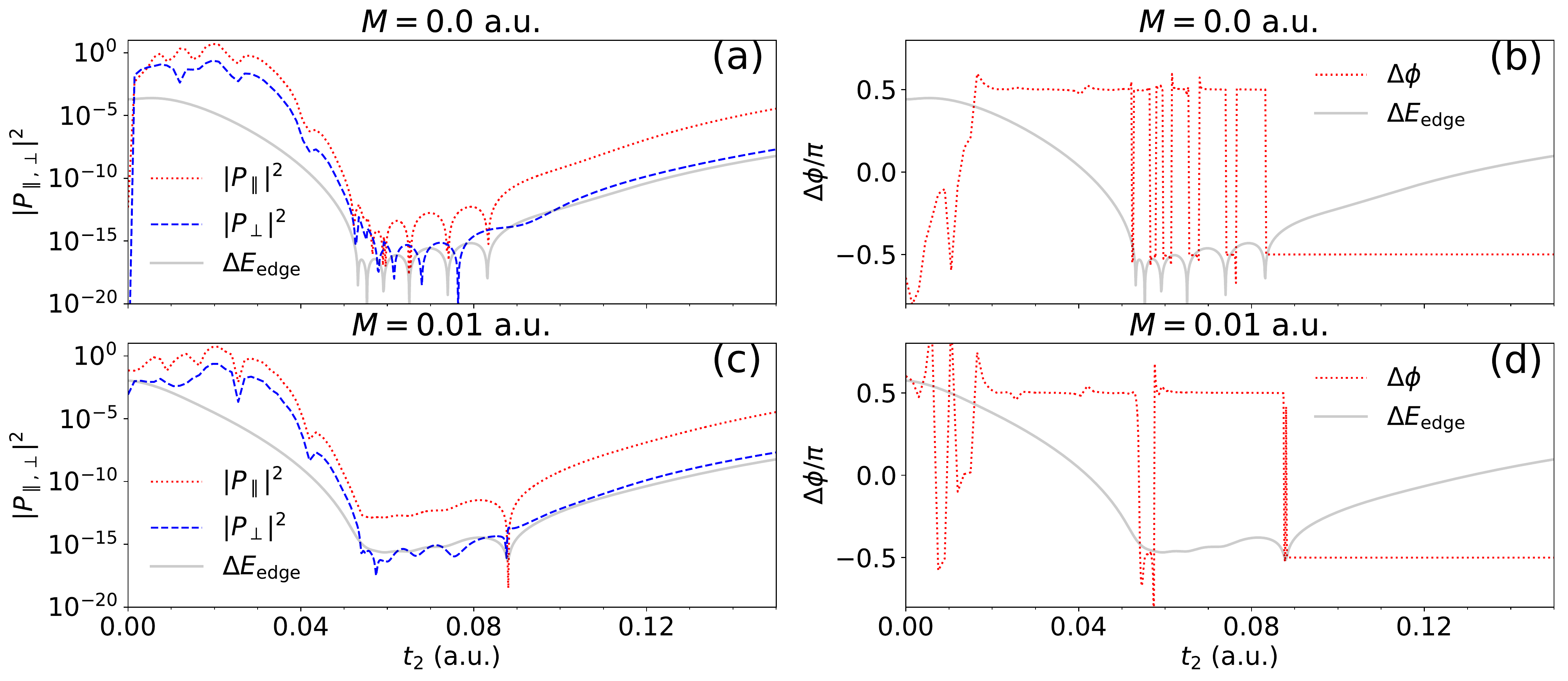}
        \caption{\label{fig7} Harmonic yield for both polarization directions (a, c) and phase difference (b, d) for harmonic order 5 as function of $t_2$. For (a, b) the on-site potential is $M = 0$, for (c, d) $M = 0.01$. The energy difference $\Delta E_\mathrm{edge}$ is included in gray to ease the interpretation of the results (extra $y$-axis for $\Delta E_\mathrm{edge}$ suppressed).}
        \end{figure*}
	   
	   Figure \ref{fig6} shows harmonic spectra in parallel (Fig.\ \ref{fig6}a) and perpendicular (Fig.\ \ref{fig6}b) polarization direction to the polarization of the incoming field for $M=0$ as function of $t_2$. In Fig.\ \ref{fig6}c, the phase difference (\ref{eq:phasediff})  between both components is shown. Just to avoid confusion: In the previous sections we discussed phases of electronic edge states $\varphi$, now we examine the phases of the emitted harmonic radiation $\Delta \phi$. The goal is to understand how both are related.

	   We only show the spectra for the parameter space where the properties of  the edge states matter. For more details at other parameters, in particular higher harmonic orders,  we refer to Ref. \cite{juerss2020helicity}. The harmonics of interest are below energy $\Delta E_\mathrm{gap}$, which is defined by the highest state of the valence band and the lowest edge state (see Fig.\ref{fig2}). The harmonic yield in this region decreases exponentially with harmonic order due to the destructive  interference of intraband harmonics \cite{bauer_high-harmonic_2018}. However, odd harmonics can still be observed up to order 9 or 11, depending on $t_2$. At certain $t_2$ the harmonic yield drops drastically for harmonics 5 till 9. This can be seen as a yellow horizontal traces in Figs.\ \ref{fig6}a,b. In the phase plot, several flips of the phase from blue to red color (flip by $\pm \pi$) can be observed. For a fixed harmonic order (5 till 9) the phase difference flips several times as $t_2$ increases.

        In Fig.\ \ref{fig7}, the harmonic yield in both polarization directions for harmonic order 5 is shown for $M = 0$ (Fig. \ref{fig7}a) and $M = 0.01$ (Fig. \ref{fig7}c) as function of $t_2$. The respective phase differences (\ref{eq:phasediff}) are plotted in Fig. \ref{fig7}b ($M = 0$) and Fig. \ref{fig7}d ($M = 0.01$). The energy difference of the edge states $\Delta E_\mathrm{edge}$ is included (with an extra $y$-axis suppressed, as only the behavior as function of $t_2$ is relevant). 
        
        In Fig.\ \ref{fig7}a, one can see a decreased harmonic yield in parallel polarization direction that occurs exactly at the points where $\Delta E_\mathrm{edge}$ is minimal (crossing of the edge states). There is an exception for the first two local minima of  $\Delta E_\mathrm{edge}$ where no significant decrease of the harmonic yield is observed. The minima of the yield in the perpendicular direction are located between two crossings (again with an exception between the first two crossings). In the phase difference (Fig.\ \ref{fig7}b) a phase flip from $\Delta \phi = \pi/2$ to $\Delta \phi = -\pi/2$ can be observed for the last four crossings. The phase flips back to $\Delta \phi = \pi/2$ between two crossing points. The back-flip of the phase is located at about the local minima of the yield in perpendicular polarization direction. Interestingly, the phase difference is $\Delta \phi = \pi/2$ before the first crossing and becomes $\Delta \phi = -\pi/2$ after the last crossing point. Further we note that the harmonic yield in parallel polarization direction is related to $\Delta E_\mathrm{edge}$ for sufficiently large $t_2$ ($t_2 > 0.055$). 
        
        In Ref.\ \cite{Silva2019} the on-site potential was finite. In Figs.\ \ref{fig7}c and \ref{fig7}d, the harmonic yield and the phase difference is  shown for $M = 0.01$. The harmonic yield in parallel polarization direction  drops again drastically at the local minimum of $\Delta E_\mathrm{edge}$ at $t_2 = 0.088$. This is the point where the energies of the edge states have an avoided crossing. The harmonic yield in perpendicular direction drops at the same value but not as much as the yield in parallel direction. Before the first local minimum of $\Delta E_\mathrm{edge}$ the phase difference is fluctuating around vales $\Delta \phi = \pi/2$. At the first local minimum at $t_2 = 0.059$ a phase flip to $\Delta \phi = -\pi/2$ can be observed but the phase flips back to $\Delta \phi = \pi/2$ for a slightly larger $t_2$. At the point of the avoided crossing at $t_2 = 0.088$, the phase flips permanently to $\Delta \phi = -\pi/2$. 
        
        The two graphs of the phase difference Figs.\ \ref{fig7}b,d show that the (avoided) crossings of the edge state energies cause a phase flip by $\pi$. Between two crossings, the phase flips back slightly after the first of the two crossings. Comparing the phases for small and large $t_2$, the phase changes from $\Delta \phi = \pi/2$ (small $t_2$) to $\Delta \phi = -\pi/2$ (large $t_2$). 
        
        The phase flips at the crossing points can be understood by the edge states. The properties of the initially occupied and unoccupied edge state exchange at each crossing point (and the avoided crossing at $t_2 = 0.088$ for $M = 0.01$). Therefore, the occupied edge state suddenly has the symmetry of the unoccupied state and the other way around. This affects the yield and the helicity of the emitted harmonics, as just demonstrated.

        In the parameter regime where the phase flips occur, the harmonic yield for both polarization directions differs several orders of magnitude for fixed $t_2$. This means that despite $\Delta \phi = \pm \pi/2$ the ellipticity of the emitted harmonics is close to zero, i.e., the harmonics are almost linearly polarized. Nevertheless, the helicity flips discussed in this work should be measurable experimentally by interferometric means.

	\section{Summary and outlook} \label{sec:summandout}

	    The edge states in the simulated finite, topological nano\-ribbons show a specific behavior as the tight-binding parameters are varied. The two edge states do not converge to the same energy but show crossings and avoided crossings. These crossings have a significant influence on the harmonic generation process. The phase difference between the two polarization components of the emitted light for certain harmonic orders change where the edge state energies cross (or have an avoided crossing). We find that the yield of low-order harmonics polarized parallel to the polarization of the incoming field  is related to the energy difference of the edge states.
	    
	    Certainly, our model studies presented in this work are highly idealized and simplified, as is the original Haldane model for the corresponding bulk. However, tailorable anomalous Hall systems are available (see, e.g.,  \cite{ChenXie2018} and references therein), and a more realistic theoretical description of HHG in such systems is  worthwhile to pursue in future work. 
	
	\begin{acknowledgements}
	C.J. acknowledges financial support by the doctoral fellowship program of the University of Rostock.
    \end{acknowledgements}

    \section*{Author contribution statement}
    C.J. performed the numerical simulations, analyzed the results, and wrote the manuscript. D.B. provided critical feedback, supported the analysis of the results, and improved the final version of the manuscript.
    
    \section*{Data Availability Statement}
    The data that support the findings of this study are available on request from the author C.J.

	\bibliography{biblio.bib}

\end{document}